\pdfoutput=1
\documentclass[10pt]{article}
\usepackage{newtxtext}
\usepackage[varg,slantedGreek]{newtxmath}
\usepackage{amsmath,amssymb,amsfonts,bm,cite}%
\usepackage{multicol,indentfirst,textcase}
\usepackage{enumerate}
\usepackage{graphicx,float}
\usepackage{capt-of}

\usepackage{ifpdf}

\ifpdf
\usepackage[bookmarks=false,pdfstartview=FitH,linkbordercolor={0.5 1 1},%
citebordercolor={0.5 1 0.5},unicode,%
hyperindex]{hyperref}
\fi
\usepackage{breakurl}

\def\BibTeX{{\rm B\kern-.05em{\sc i\kern-.025em b}\kern-.08em
    T\kern-.1667em\lower.7ex\hbox{E}\kern-.125emX}}

\date{}

\begin{document}

\title{\bf Caution, DOI!\\
\medskip\Large Bibliographic detective story in the era of
digitalization}

\author{Victor Kozyakin\\
\small Kharkevich Institute for Information Transmission Problems\\
\small Russian Academy of Sciences\\
\small Bolshoj Karetny lane 19, Moscow 127051, Russia}

\maketitle
\begin{abstract}
An example of inconsistencies in information provided by popular
bibliographic services is described and the reasons for these
inconsistencies are discussed.\\

\indent\textbf{Keywords:} bibliography, digital object identifier, DOI,
BibTeX, RIS
\end{abstract}
\section*{}
\hfill\parbox{7cm}{\it\flushright\small The best laid schemes o' Mice an' Men
Gang aft agley.\\ Robert Burns, ``To a Mouse''}~\bigskip
\begin{multicols}{2}

\section*{Introduction}
The collection of exact and complete bibliographic references is inevitable
in scientific research. Researchers of the precomputer era remember how
difficult it was to collect a reference list on paper correctly and in
accordance with the requirements for journal reference list typography. The
situation changes drastically as computer methods for publications
preparation have been introduced---we will only talk about the \LaTeX{}
system and its various add-ons, although this is, to a certain extent, true
for other typography software systems, both proprietary and free.

The \BibTeX{} software created by Oren Patashnik became a widespread and,
importantly, convenient way of preparing bibliographic references. The
process of preparing bibliographic references with \BibTeX{} is divided into
two stages: the manual preparation of the database of the needed references,
in which for every publication we typeset the necessary bibliographic data in
a certain format and the automatic (with \BibTeX{}) typography of the
reference list according to \texttt{.bst} style files that are designed by
many publishers following bibliographic typography and citation styling
preferences.

Note that the creation of a database of publications for subsequent
processing with \BibTeX{} takes time and requires attention and certain
knowledge about the rules for formatting its structural elements. However,
the time and effort it takes to typeset the publication database is more than
compensated by the simplicity of the next application of these databases in
formatting bibliographies in different papers and, more importantly, by the
crucial reduction of errors in publication formatting. Note also that the
manual composition of the \BibTeX{} database is not necessary in most cases,
as a rule, because the needed records are usually formed by publishers and
many bibliographic online services in the required format.

\begin{figure*}\centering
\includegraphics[width=0.6\paperwidth,clip]{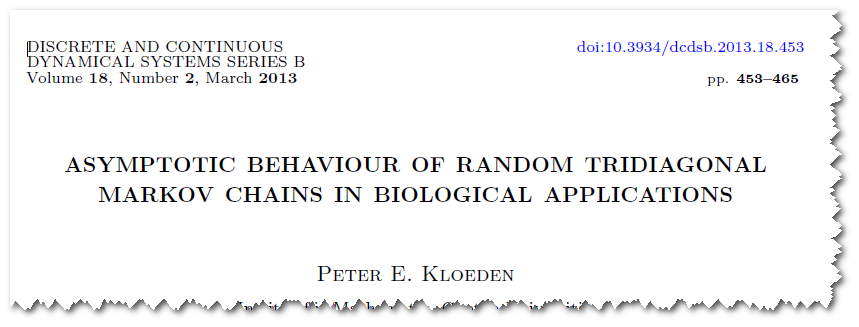}
\caption{Fragment of title page with publisher's imprint of paper~\cite{KK:DCDSB13}}\label{F:1}
\end{figure*}

Figure~\ref{F:1} shows a fragment of the title page with the publisher's
imprint of a paper~\cite{KK:DCDSB13}.

Below, we present the record in \BibTeX{} format corresponding to this paper,
from the bibliographic system \textsc{MR Lookup} of the American Mathematical
Society
(\href{https://mathinet.ams.org/mrlookup}{https://mathscinet.ams.org/mrlookup}):

\noindent{\scriptsize\begin{verbatim} @article {MR2999086,
    AUTHOR = {Kloeden, Peter E. and Kozyakin, Victor S.},
     TITLE = {Asymptotic behaviour of random tridiagonal
              {M}arkov chains in biological applications},
   JOURNAL = {Discrete Contin. Dyn. Syst. Ser. B},
  FJOURNAL = {Discrete and Continuous Dynamical Systems.
              Series B. A Journal Bridging Mathematics
              and Sciences},
    VOLUME = {18},
      YEAR = {2013},
    NUMBER = {2},
     PAGES = {453--465},
      ISSN = {1531-3492},
   MRCLASS = {60J10 (15B48 92C99)},
  MRNUMBER = {2999086},
MRREVIEWER = {Ross S. McVinish},
       DOI = {10.3934/dcdsb.2013.18.453},
       URL = {https://doi.org/10.3934/dcdsb.2013.18.453}}
\end{verbatim}}
\newpage
A similar record provided from the \mbox{\textsc{zbMATH}} system of the
European Mathematical Society
(\href{https://zbmath.org}{\texttt{https://zbmath.org}}) is as follows:

\noindent{\scriptsize\begin{verbatim} @Article{zbMATH06146721,
    Author = {Peter E. {Kloeden} and Victor {Kozyakin}},
    Title = {{Asymptotic behaviour of random tridiagonal
             Markov chains in biological applications.}},
    FJournal = {{Discrete and Continuous Dynamical Systems.
                Series B}},
    Journal = {{Discrete Contin. Dyn. Syst., Ser. B}},
    ISSN = {1531-3492; 1553-524X/e},
    Volume = {18},
    Number = {2},
    Pages = {453--465},
    Year = {2013},
    Publisher = {American Institute of Mathematical Sciences
                 (AIMS), Springfield, MO},
    Language = {English},
    MSC2010 = {60J10 15B48 37H10},
    Zbl = {1277.60118}}
\end{verbatim}}

We see that all significant bibliographic information (authors, name of the
journal, publisher's imprint, etc.) in both \BibTeX{} records coincides. At
the same time, the formatting of the corresponding records is slightly
different, and, in addition, include some individual fields (for instance,
the identification numbers in the corresponding systems: \texttt{MRNUMBER}
and \texttt{Zbl}) reflecting the preferences of the authors of these records.
In particular, the record of the \textsc{MR Lookup} contains an important
\texttt{DOI} field, which is the digital object identifier, with which we can
jump to the publisher's webpage, at least containing the annotation and
bibliographic data of the sought publication (sometimes, its full text),
using the International DOI Foundation (IDF) service
(\href{http://www.doi.org}{\texttt{http://www.doi.org}}).

\section*{Finally, we have arrived at DOI}

The DOI system was created due to publishing industry initiative, which
admits the need for unique identification of the content objects, rather than
reference to their location. In 1998, the International DOI Foundation was
founded to develop the system; the necessary technologies and standards have
been created for the introduction of the DOI
system~\cite{DOIHandbook,Paskin09}. The first service for registering the DOI
names began operating in 2000, and, towards the beginning of 2009, there were
already allocated approximately eight million DOI names through eight
registration services. The most used application of the DOI system is the
service of cross links between publishers called \textsc{Crossref}
(\href{https://www.crossref.org}{https://www.crossref.org}), which allows
associating references from the citation directly with the cited content on
the platform of another publisher with account for the access-control methods
of the goal publisher.

The original DOI names may be represented by long lines of symbols, which is
sometimes inconvenient for reference organization. To avoid this, the
International DOI Foundation opened a service of the reduced DOI names called
\textsc{shortDOI} (\href{http://shortdoi.org}{http://shortdoi.org}). When we
request the \textsc{shortDOI$^\circledR$} with the original DOI, its
shortened nickname is created in the format \texttt{10/abcde} (or the
previously shortened nickname is returned), and we may work with it further
as with the original DOI.

\begin{figure*}\centering
\includegraphics[width=0.66\paperwidth,clip]{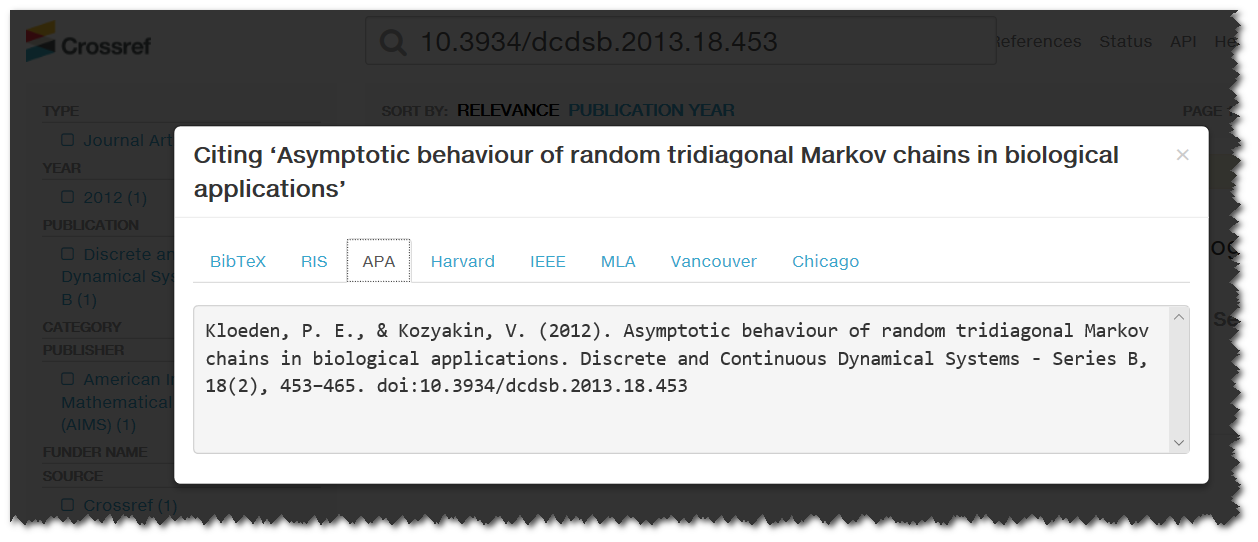}
\caption{Screenshot of \textsc{Crossref} service response on request
DOI 10.3934/dcdsb.2013.18.453}\label{F:2}
\end{figure*}

Currently, various functions with the use of DOI are carried out by multiple
services, proprietary and free bibliography managers, such as
\begin{itemize}
\item \textsc{EndNote} by the Clarivate Analytics company,\\
    \href{https://endnote.com}{https://endnote.com},
\item \textsc{Mendeley} by the Elsevier company,\\
    \href{https://www.mendeley.com}{https://www.mendeley.com},
\item \textsc{Citavi} by the Swiss Academic Software company,\\
    \href{https://www.citavi.com}{https://www.citavi.com},
\item \textsc{Zotero}, a free and open-source reference management
    software, \href{https://www.zotero.org}{https://www.zotero.org},
\item \textsc{ZoteroBib}, \href{https://zbib.org}{https://zbib.org} (a
    simplified variant of Zotero),
\item \textsc{Docear}, \href{http://www.docear.org}{http://www.docear.org},
\end{itemize}
free desktop applications
\begin{itemize}
\item \textsc{JabRef}, \href{http://www.jabref.org}{http://www.jabref.org}
    (Windows, Linux, \mbox{MacOS}),
    \item \textsc{BibDesk},
        \href{https://bibdesk.sourceforge.io}{https://bibdesk.sourceforge.io}
        (MacOS),
\item \textsc{KBibTeX},
    \href{https://userbase.kde.org/KBibTeX}{https://userbase.kde.org/KBibTeX}
    (Linux),
\end{itemize}
as well as many other internet services and desktop applications among which
we also name the \textsc{doi2bib} service
(\href{https://www.doi2bib.org}{https://www.doi2bib.org}), converting the DOI
names into the bibliographic records in \BibTeX{} format.

The introduction of the DOI system dramatically changes the entire technology
of using bibliographic data---the users become a tool for instantaneous
access to the electronic version of a publication through the \textsc{Digital
Object Identifier} service and for the same instantaneous access to the
required bibliographic information using the above-mentioned
\textsc{Crossref}, \textsc{EndNote}, \textsc{Mendeley} services, etc.

This seems a time for the universal happiness of bibliographic data users,
when all required data may be practically instantaneously obtained, having
been verified. However, this appears unfortunately not a blessing (see the
epigraph).
\begin{enumerate}
  \item Dead DOIs have appeared that correspond to nothing. The reasons may
      be various: an error in the DOI name, a closing and structural change
      of the website where the corresponding publication was located, the
      transfer of the publication to another website, etc.
  \item Semidead DOIs have appeared that are processed by some services,
      but ignored by others. For instance, the DOI \texttt{10.1000/182} of
      publication~\cite{DOIHandbook} is apparently processed only by the
      \textsc{Digital Object Identifier} service and is not processed by
      the \textsc{Crossref}, \textsc{EndNote} and \textsc{Mendeley}
      services. The situation in which the DOIs generated by the request of
      the users of researchers' social network \textsc{ResearchGate}
      (\href{https://www.researchgate.net}{https://www.researchgate.net})
      is the same..
  \item The wrong DOIs that point to other publications.
  \item Finally, the bibliographic data provided by different services on
      the DOI request may differ. For instance, the data
      request~\cite{KK:DCDSB13} with DOI \texttt{10.3934/dcdsb.2013.18.453}
      to the \textsc{Crossref} service leads to the citation given in
      Fig.~\ref{F:2}, where the 2012 year of publication is different from
      the true year (2013) of the journal (print) publication. A similar
      situation appears upon request from the \textsc{EndNote},
      \textsc{Mendeley}, \textsc {ZoteroBib}, \textsc{doi2bib} services and
      on attempting to obtain data via \textsc{JabRef}---all provide the
      wrong year of publication of paper~\cite{KK:DCDSB13} with its DOI.
\end{enumerate}

The first two mentioned disadvantages are not critical. Here, at least, the
corresponding service requested for bibliographic data on the DOI directly
informs us that these data cannot be given. The third disadvantage causes
annoyance, however everyone can make a mistake. Fortunately, the first three
disadvantages have an accidental character.

The last disadvantage appears to be sufficiently unpleasant because it
manifests systematically and none of the above-mentioned services inform us
that the data given require additional verification. This lowers the sense of
the digital object identifier, the DOI, to a large extent.

\section*{Investigation}

A reasonable question arises: How could it be that different services provide
different information on the same DOI?

We note that \BibTeX{} is not the only nor most widespread format for storing
bibliographic data. \BibTeX{} became widely used in the scientific
publications environment prepared mostly with the \LaTeX{} system and its
various add-ons. In the publishing industry, the different formats for
storing and exchanging bibliographic data are the most widespread (appearing
well before \BibTeX{}).Among them one of the most widely used is the RIS
format developed by Research Information System and applied as the main
format of digital libraries such as \textsc{IEEE Xplore}, \textsc{Scopus},
\textsc{ScienceDirect}, and \textsc{SpringerLink} and bibliographic services
such as \textsc{Zotero}, \textsc{Citavi}, \textsc{Mendeley},
\textsc{EndNote}, and \textsc{Crossref}. The record for
paper~\cite{KK:DCDSB13} in RIS from the \textsc{Crossref} has the following
form:

\noindent{\scriptsize\begin{verbatim}
 TY  - JOUR
 DO  - 10.3934/dcdsb.2013.18.453
 UR  - http://dx.doi.org/10.3934/dcdsb.2013.18.453
 TI  - Asymptotic behaviour of random tridiagonal Markov
       chains in biological applications
 T2  - Discrete and Continuous Dynamical Systems - Series B
 AU  - Kloeden, Peter E.
 AU  - Kozyakin, Victor
 PY  - 2012
 DA  - 2012/11
 PB  - American Institute of Mathematical Sciences (AIMS)
 SP  - 453-465
 IS  - 2
 VL  - 18
 SN  - 1531-3492
 ER  -
\end{verbatim}}
\noindent In this record there are two parameters characterizing the data:
\texttt{DA (Date)} and \texttt{PY (Publication year)}, and in both the
\textbf{2012} year is specified!

Unfortunately, in the RIS description available via

\noindent\href{https://web.archive.org/web/20120526103719/http://refman.com/support/risformat_intro.asp}{https://web.archive.org/web/20120526103719\\
/http://refman.com/support/risformat\_intro.asp},

\noindent there is no detailed explanation on the sense of these parameters.
However, the value \texttt{2012/11} of the \texttt{PY} parameter in the RIS
record of paper~\cite{KK:DCDSB13} coincides with the publication of the first
online version of this paper. This is probably the sense of the \texttt{PY}:
parameter: the year of the first public appearance of the publication. At the
same time, the \texttt{Year} parameter in \BibTeX{} is described as the
imprint year of the printed publication. Apparently, it is the reason, i.e.,
the never mentioned difference in the interpretation of the concept
``publication year'', that leads to inconsistencies in the bibliographic data
of paper~\cite{KK:DCDSB13} provided by the \textsc{MR Lookup} and
\textsc{zbMATH} on the one side and \textsc{Crossref}, \textsc{Mendeley},
\textsc{ZoteroBib}, \textsc{doi2bib}, \textsc{JabRef} on the other side.

\section*{Conclusions}

We have appealed by describing inconsistency in the data provided by
different bibliographic services to the forum of an application (its name is
insignificant) unambiguously promoting the idea of the advantage of receiving
bibliographic data from internet sources. Unfortunately, this appeal led to
nothing: we received an answer ``explaining'' that the application has
nothing to do with it and relies on data provided by the services without
verification. Organizing the dialog between the two groups of information
services with an invitation to coordinate/standardize ways of interpreting
the bibliographic data is more than an ordinary user can manage. And, taking
into account that the situation described above is not unique, we conclude
this note: Do you use DOI? Trust but verify.


\flushright {\emph {Translated by E. Oborin}}
\end{multicols}

\end{document}